\def\AaA#1#2{{A\&A} #1\rm, #2}

\def\ApJ#1#2{{ApJ} #1\rm, #2}

\def\nat#1#2{{Nat} #1\rm, #2}

\input psfig
  \MAINTITLE={Illumination in binaries}
  \FOOTNOTE{Tables 1 and 2 are only available in electronic form at the CDS 
            via anonymous ftp 130.79.128.5}
  \SUBTITLE={ ????? }
  \AUTHOR={J.-M. Hameury@1, H. Ritter@2}
  \OFFPRINTS={J.-M. Hameury}
  \INSTITUTE={ @1 URA 1280 du CNRS; Observatoire de Strasbourg, 
                  11 rue de l'Universit\'e, F-67000 Strasbourg, France
               @2  Max-Plank-Institut f\"ur Astrophysik,
                  Karl-Schwarzschild-Str. 1, 85740 Garching, Germany}
  \DATE={ ????? }
  \ABSTRACT={We give a simple, but accurate method that can be used to
account for illumination in compact binary systems which have a low-mass
companion, even if spherically symmetric illumination of the secondary star
(not necessarily on the main sequence) is not assumed. This is done by
introducing a multiplicative factor $\Phi$ in the Stefan-Boltzmann surface
boundary condition, which accounts for the blocking of the intrinsic
secondary flux by X-ray heating of the photospheric layers.  Numerical fits
and tables for $\Phi$ are given for unperturbed effective temperatures in
the range 2500 -- 5600 K and $\log g$ in the range 1.0 -- 5.0}
  \KEYWORDS={accretion, accretion discs --  (Stars: ) binaries: general --
             X-rays: stars}
  \THESAURUS={ 02.01.2;  
               08.02.1;  
               13.25.5)  
                       }
\maketitle
\titlea{Introduction}

The effects of irradiation of the low mass companion in a low-mass X-ray
binary or a cataclysmic variable have received much attention in the recent
years. In cataclysmic variables, the external illumination flux due to
accretion onto the primary compact object is comparable to the intrinsic
stellar flux produced by nuclear reactions, whereas this external flux
exceeds by orders of magnitude that of the secondary in low-mass X-ray
binaries. A fraction of the illumination flux is absorbed in optically thin
regions of the secondary photosphere, and this may result in the formation of
a wind (see e.g. London et al. 1981; Ruderman et al. 1989; Tavani \& London
1993), which can in turn affect the long term evolution of the system. We
shall not discuss this effect here, but instead consider energy deposition
below the photosphere of the secondary.

Early calculations in the case of low-mass X-ray binaries (Podsiadlowski
1991; Harpaz \& Rappaport 1991) in which it was assumed that the secondary is
illuminated in a spherically symmetric manner showed a quite dramatic effect.
The intrinsic luminosity of the secondary star is very efficiently blocked by
the impinging radiation in the outer convective layers, so that the secondary
must expand until it becomes almost fully radiative. The secular evolution is
then quite different from the standard case, namely the systems rapidly
evolve towards longer periods, with mass transfer rates close to or above
the Eddington limit. This would have accounted for the fact that the period
distribution of LMXBs seems to be significantly different from that of
cataclysmic variables, showing a lack of systems at short periods. This would
have also increased by a large factor the number of LMXBs, making it
compatible with what is required to account for the observed density of
millisecond pulsars which are believed to be the descendants of LMXBs (Frank
et al. 1992). However, the spherically symmetric assumption is obviously
incorrect, and it was later shown by Hameury et al. (1993) and confirmed by
Harpaz \& Rappaport (1995) that the unilluminated side can efficiently cool
the secondary. The secular evolution of LMXBs is not drastically different
from the unilluminated case, although the short term behaviour is very
significantly affected by illumination, and exhibits on and off states, with
relatively high mass transfer rates.

In the simpler context of cataclysmic variables, Ritter et al. (1995, 1996a)
showed that irradiation-induced mass transfer cycles could also be present in
these systems; King (1995) and King et al. (1995, 1996) discussed from a very
general point of view the occurrence of such mass transfer cycles in CVs. The
existence of these cycles could be responsible for the observed spreading of
mass transfer rates $\dot{M}$ for a given orbital period $P$, whereas models
would predict a fair correlation between $\dot{M}$ and $P$. Ritter et al.
(1996b) showed that, in the bi-polytrope approximation (Kolb \& Ritter 1992),
the stability criterion depends critically upon the variation of the
effective temperature of the illuminated star as a function of the
irradiating flux. For modeling the response of the stellar surface to
illumination, they used a very simple one zone model for the
superadiabatic layers of the low-mass secondary. Here, we use detailed
stellar models to calculate the response of the secondary, namely the
secondary luminosity as a function of irradiation, for a range of values of
the surface gravity and unperturbed effective temperature. We give our
results in a tabular form that can be used to calculate the evolution of a
compact system in the presence of illumination. The advantage of this
formulation is that it can be used in bi-polytropic codes, which are much
faster that full stellar codes, and thus allow the exploration of a wide
range of parameters, but have to be calibrated.

\titlea{Reaction of the stellar surface to external irradiation}

The influence of heating of subphotospheric layers by an external irradiation
flux is entirely described by a modification of the boundary condition, as
all of the absorbed X-ray flux is thermalized before being re-emitted. In the
plane parallel approximation, the standard Stefan-Boltzmann law $L = 4 \pi
R_2^2 \sigma T_{\rm eff,0}^4$ where $L$ is the surface luminosity of the
secondary star in the absence of illumination (note that $L$ need not be
equal to the nuclear luminosity), $R_2$ its radius, and $T_{\rm eff,0}$ is
the effective temperature of the unilluminated star has to be replaced by
(Ritter et al. 1996b):
$$
L = R_2^2 \sigma T_{\rm eff,0}^4 \int_0^{2\pi}\int_0^{\pi} G(x(\theta,\phi))
\sin \theta d\theta d\phi \; , \eqno\autnum
$$
where $x = F_{\rm irr}/\sigma T_{\rm eff,0}^4$ is the normalized
illuminating flux, and  
$$ 
G(x)= \left( {T_{\rm eff} (x) \over T_{\rm eff,0}} \right)^4 -x\; .
\eqno\autnum
$$
Here $T_{\rm eff}$ is the effective temperature of a stellar surface element
subject to the illuminating flux $x$.
The function $G(x)$, in the range 0 -- 1, describes the blocking of the
intrinsic stellar luminosity as a result of illumination. $F_{\rm irr}$
includes only the fraction of the flux that is deposited below the
photosphere; as mentioned above, energy deposition in optically thin regions
does not result in modifications of the internal structure of the star. 

In the case of low-mass secondaries which have a convective envelope, the
entropy deep in the envelope remains constant over the whole surface of the
star, and varies slowly with time, on the Kelvin-Helmholtz time of the whole
envelope. By contrast, the superadiabatic and radiative outer layers have a
very short thermal time, and are in thermal equilibrium. Moreover, Gontikakis
\& Hameury (1993) have shown that the flux, in the plane parallel
approximation, does not vary with depth in the convective zone. The problem of
determining the reaction of the star to illumination thus reduces to finding
the structure of an initially convective layer in thermal equilibrium when
one changes the outer boundary condition while keeping the entropy at its
base constant. This is easily done solving the standard stellar structure
equations, which, in the plane parallel approximation write:
$$
\eqalign{
        & {d \log P \over d \log \Sigma } = g \cr
        & {d \log T \over d \log P } = \nabla  \; , \cr
        } \eqno\autnum
$$

\begfig 0.00cm
\psfig{figure=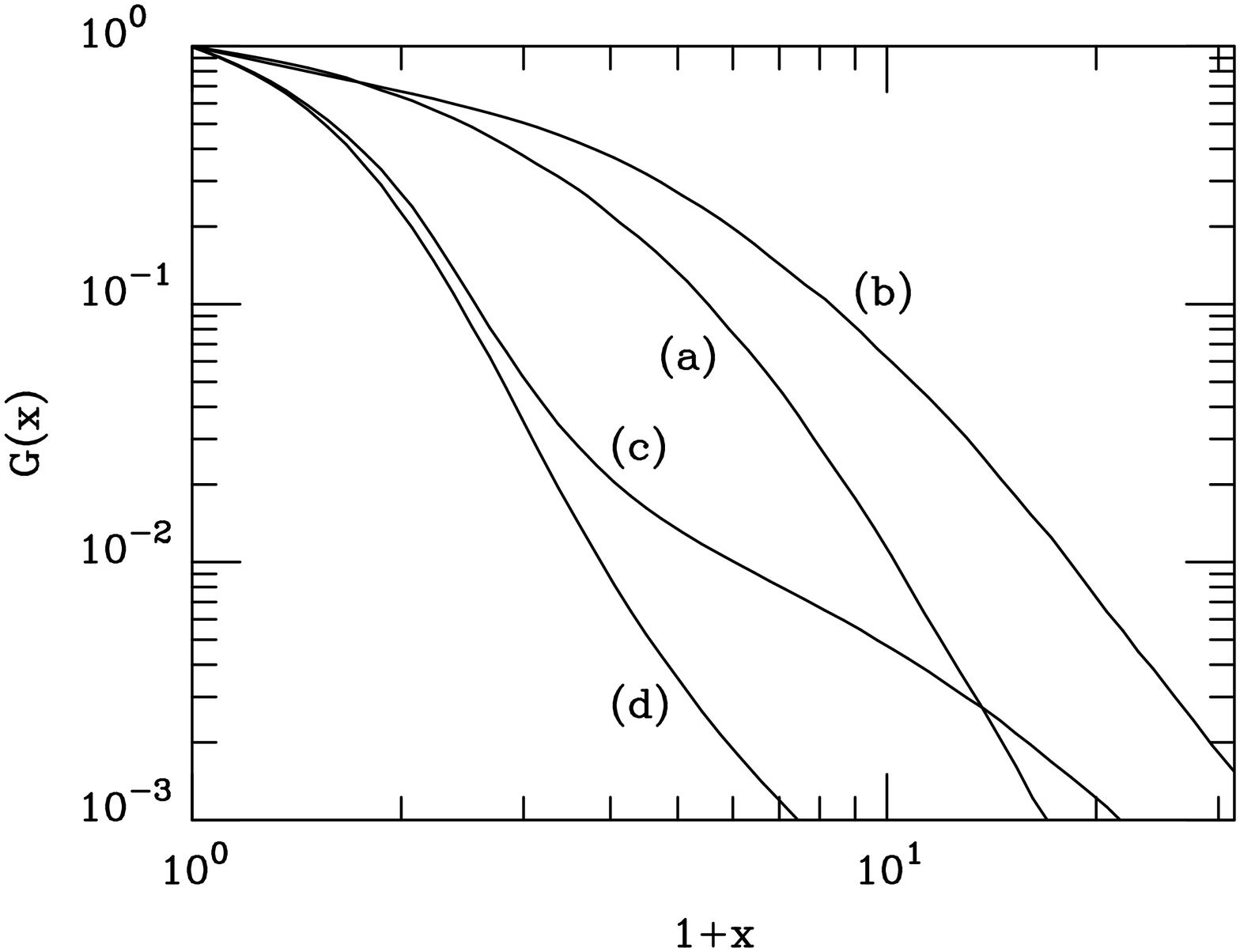,width=8.75cm}
\figure{1}{Function $G(x)$ for secondaries on the main sequence, with a mass
0.8 (a), 0.6 (b), 0.4 (c) and 0.2 M$_\odot$ (d)}
\endfig

\begfig 0.00cm
\psfig{figure=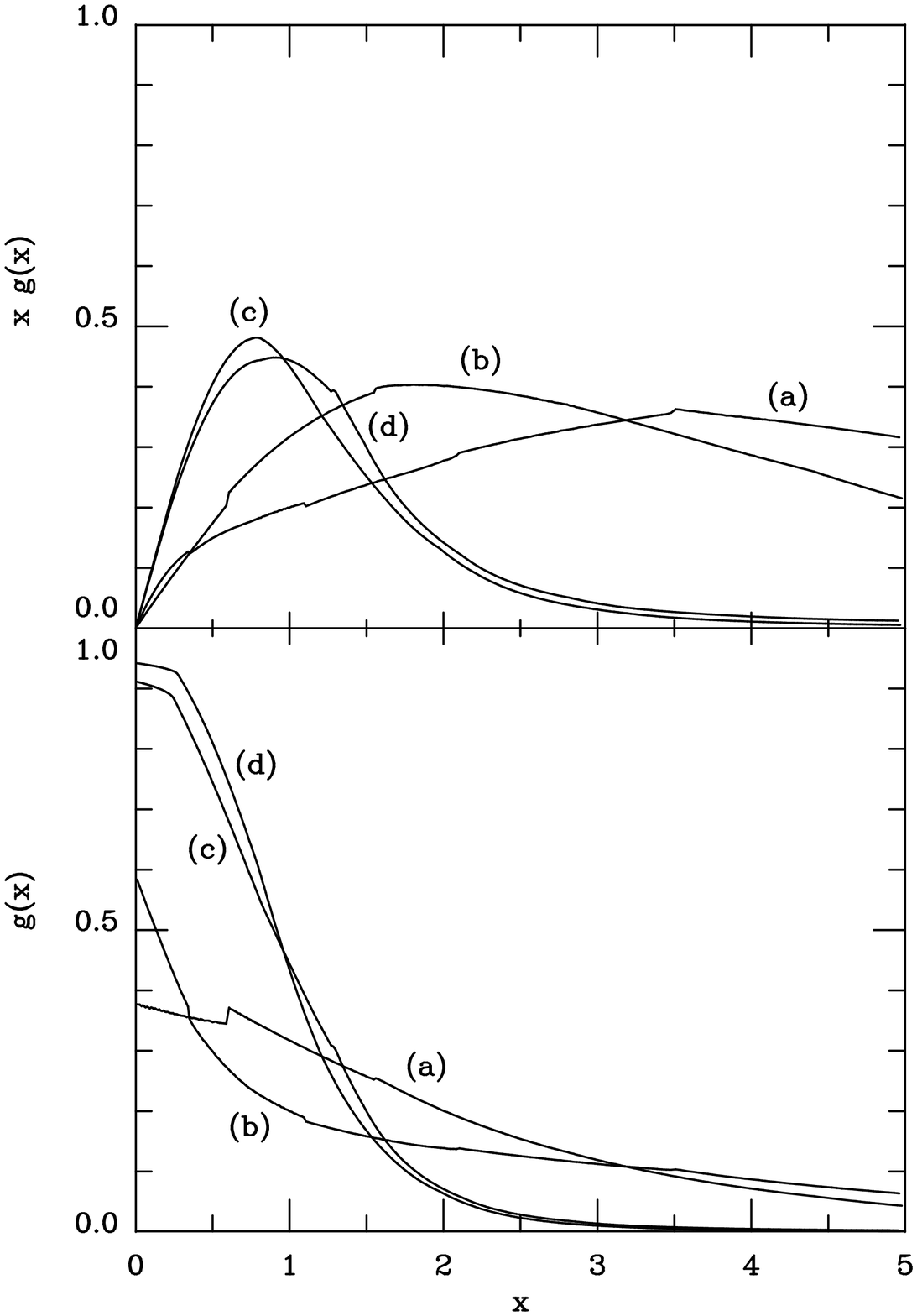,width=8.75cm}
\figure{2}{Functions $g(x) = - dG(x)/dx$ and $x g(x)$ for secondaries on the
main sequence with masses 0.8 (a), 0.6 (b), 0.4 (c) and 0.2 M$_\odot$ (d).}
\endfig

\noindent where $P$ is the pressure, $T$ the temperature, $\nabla$ the
temperature gradient calculated using the mixing length approximation in the
convective zone, $\Sigma$ the column density, and $g$ the surface gravity,
assumed to be constant. The energy flux $F$ is also assumed to be constant
throughout the layer. The layer is integrated down to a depth $\Sigma_{\rm
max}$ of $10^6$ g cm$^{-2}$, which is deep enough that the departure from
adiabaticity is negligible; our results have been found to be independent of
the particular value of $\Sigma_{\rm max}$. The surface boundary conditions
are standard:
$$
\kappa P = 2/3 g, \quad F + F_{\rm irr}= \sigma T^4 \eqno\autnum
$$
In the absence of illumination, these equations are integrated for a given
$g$ and $T_{\rm eff}$, and give the entropy $S_0$ deep in the convective zone.
In the presence of an irradiating flux, one adds the condition $S = S_0$ at
the base of this layer. The set of equations (3) are integrated using the
method, equation of state and opacities described in Hameury (1991). This
gives $F$, and thus $G$ as a function of $g$, $T_{\rm eff,0}$, and $x$.
Figure 1 shows $G(x)$ for main sequence stars of various masses. A
significant difference can be seen between very low mass secondaries (less
than 0.4 M$_\odot$) and more massive ones. This difference is due to the low
value of the energy flux that has to be carried through the convective zone,
and hence to a small deviation to adiabaticity, even in the outer
superadiabatic layers. For those low-mass stars, $G(x)$ is not very
different from the relation $G(x) = \max(1-x,0)$ that one would obtain
assuming that there is no superadiabatic region, so that the temperature
gradient and thus the surface temperature is fixed.

The stability analysis of mass transfer involves the derivative of the
function $G(x)$ (Ritter et al. 1996c), $g(x) = - dG(x)/dx$. More precisely,
the criterion for stability against irradiation-induced mass transfer is:
$$
C = 2 (\zeta_{\rm s} - \zeta_{\rm R}) {\tau_{\rm KH,0} \over {\tau_{\rm
\dot{M}}}} {\cal F}^{-1} > 2 s x g(x) \eqno \autnum
$$
where $\zeta_{\rm s}$ is the adiabatic mass radius exponent of the
secondary, $\tau_{\rm KH,0}$ the Kelvin-Helmholtz time, $\tau_{\rm\dot{M}}$
the mass transfer time scale, $\cal F$ a dimensioneless function which scales
roughly as the inverse of the mass of the secondary convective envelope, $s
\sim 1/2$ the fraction of the secondary exposed to illumination, and
$\zeta_{\rm R}$ is the mass radius exponent of the Roche radius. This
criterion is deduced from the assumption that the mass-radius exponent of the
secondary under the sole influence of illumination must be less than
$\zeta_{\rm s} - \zeta_{\rm R}$; Eq. (5) means that, for instability to
occur, one must satisfy both conditions of sufficient illuminating fluxes
and sensitivity of the secondary response to irradiation, i.e. large value of
$g(x)$.
$C$ does
not depend on the very detailed structure of the illuminated star, and is
easily accessible in the bi-polytropic approximation. The response of the
secondary is entirely contained in the function $x g(x)$, which is plotted in
Fig. 2. The small discontinuities are due to linear interpolations in
determining the opacity.  A remarkable characteristic of this function is
that (1) $g(x)$ is a monotonously decreasing function , which is less than
unity, and (2) that $x g(x)$ has a maximum of the order of 0.5, whatever the
secondary mass, although the position of this maximum does depend on it.
Thus, there will be no cycles if $C > s$ (note that this is a sufficient but
not necessary condition for stability).

It must be stressed that this procedure is valid only for stars which have a
convective envelope, even though the whole star can be far out of thermal
equilibrium. This would in fact be the main limitation in following the
evolution of systems in which the effect of irradiation is so strong that at
some point the star becomes fully radiative. This was a natural outcome of
models assuming spherically symmetric illumination, but Hameury et al. (1991)
have shown that this is no longer the case when one accounts for the
asymmetry of irradiation.

These calculations can easily be generalized to stars which are not on the
main sequence, provided they still have a convective envelope. Given the
effective temperature $T_{\rm eff}$ and surface gravity $g$, the set of
equations (3) can be integrated in the absence of illumination and provide
the unperturbed outer structure; the same procedure as previously is then
applied. Table 1 (available only in electronic form at the CDS via anonymous
ftp 130.79.128.5) gives $-\log_{10} (G)$
(column 4) as a function of $\log g$ (column 1), $\log T_{\rm eff}$ (column
2) and $\log (1+x)$ (column 3).

\begfig 0.00cm
\psfig{figure=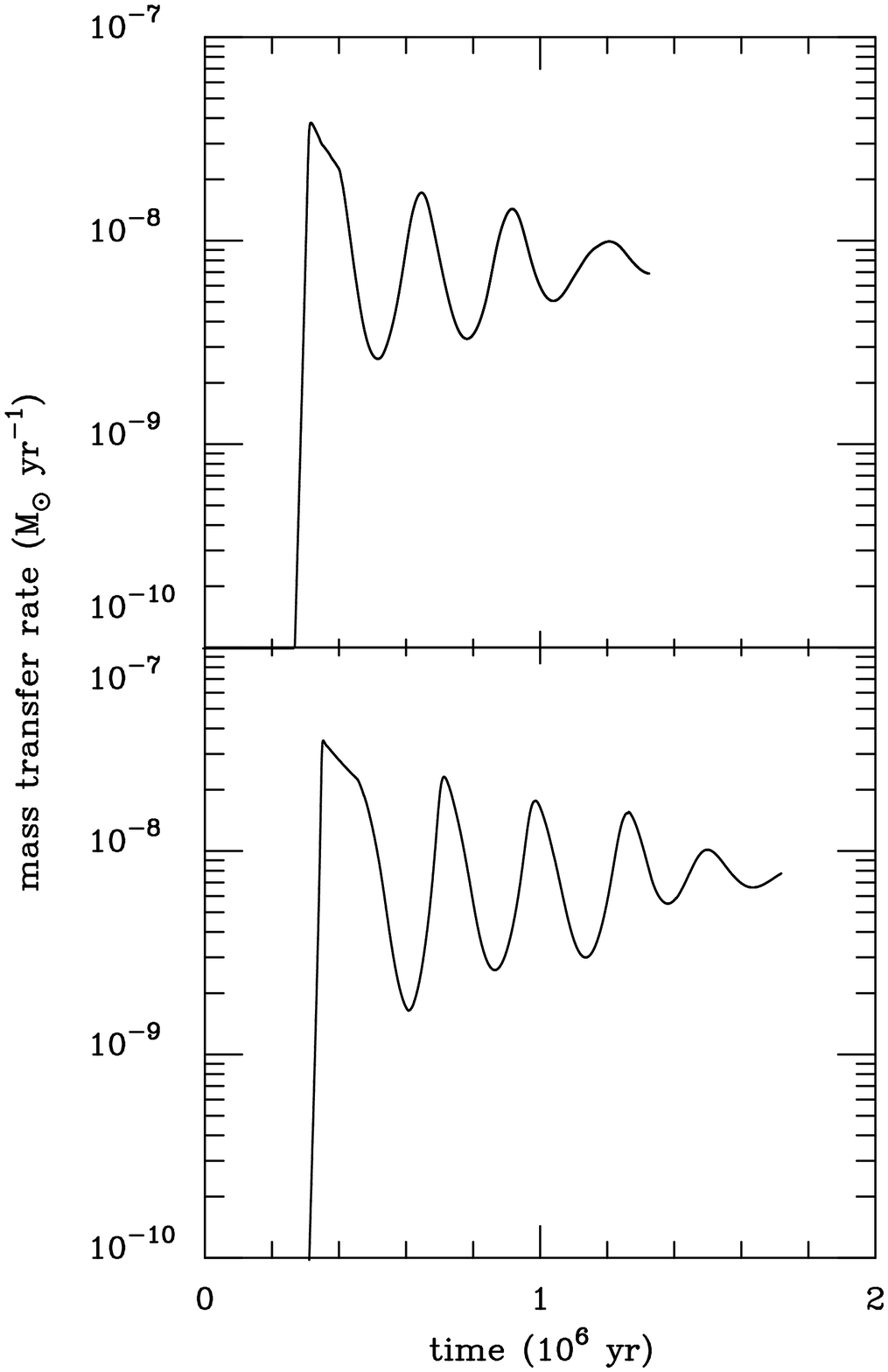,width=8.75cm}
\figure{3}{Evolution of a cataclysmic variable using either a two
hemispheres model with standard boundary conditions (top panel), or a
spherically symmetric model with condition (9)}
\endfig 

If one assumes that the irradiation source is a point located at the distance
$a$ from the center of the secondary, Eq. (1) can be easily integrated over
the whole stellar surface, which gives, following the notations of Ritter et
al. (1996b):
$$
L =2 \pi R_2^2 \sigma T_{\rm eff,0}^4 \left[ 1 + f_2(q) + 
\int_0^{\theta_{\rm max}} G(x(\theta)) \sin \theta d \theta \right] ,
\eqno\autnum
$$
where $\theta$ is the colatitude with respect to the substellar point of a
point on the secondary surface, $f_2(q) = R_2/a$, with $q= M_2/M_1$ the mass
ratio, and the normalized irradiation flux $x$ is now given by:
$$
x(\theta)= x_0 h(\theta) \eqno\autnum
$$
with 
$$
x_0  = {\eta \over 4 \pi} {G M_1\dot{M} \over R_1 a^2} /\sigma T_{\rm eff,0}^4
\; , \eqno\autnum
$$
where $\eta$ is an efficiency factor accounting for the albedo of the
secondary, the fraction of energy deposited in optically thin regions or the
anisotropy of emission at the surface of the primary, and
$$ h(\theta) = {\cos \theta -f_2 \over (1-2 f_2 \cos \theta +f_2^2)^{3/2} }
\; . \eqno\autnum
$$
Equation (6) can thus be written in the form
$$
L = 4 \pi R_2^2 \sigma T_{\rm eff,0}^4 \Phi (x_0) \; . \eqno\autnum
$$
We have integrated numerically Eq. (6), and the results can be fitted within
1 \% by:
$$ \eqalign{
& y = \ln (1+x_0) \cr
& \Phi (x_0) = {1 +f_2 \over 2} + {1 - f_2 \over 2} \exp (a_1 y +a_2 y^2
+a_3 y^3) \cr
           } \eqno\autnum $$
Table 2 (available only in electronic form at the CDS via anonymous ftp
130.79.128.5) gives the coefficients $a_1$,
$a_2$, and $a_3$ as a function of $\log T_{\rm eff,0}$ and $\log g$. For
arbitrary values of $\log T_{\rm eff,0}$ and $\log g$, a linear interpolation
of $y$ (not of the coefficients $a_1$, $a_2$ and $a_3$) can be done.

The stability criterion now involves the integral of the function $x g(x)$;
mass transfer will be stable if:
$$
C > \max \int_0^{\theta_{\rm max}} x(\theta) g(x(\theta)) \sin \theta d\theta
\eqno \autnum
$$
We have calculated this maximum using the values of $G(x)$ as determined
above, and found that this maximum is again almost independent of the
secondary mass, and is smaller by about a factor 2 than the value $2s$ found
assuming uniform illumination over a fraction $s$ of the secondary, being
equal to (0.2 -- 0.3) $(1-f_2)$.

We have compared evolutionary calculations using either a spherically
symmetric code with a boundary condition given by (10), or the code described
in Hameury et al. (1993) in which both hemispheres with a coupling term are
modeled. The results are given in Fig. 3, for a cataclysmic variable with
$M_1$ = 1 M$_\odot$ and $M_2$ = 0.8 M$_\odot$; the white dwarf radius is $5
\times 10^8$ cm, and the efficiency $\eta$ is 0.1 (spherically symmetric
model) and 0.09 (two hemispheres model). Both are very similar, i.e. the mass
transfer rate exhibits damped oscillations; the different values of $\eta$
were chosen so as to obtain the same initial maximum value of the mass
transfer rate. The differences (slightly shorter damping time and $\eta$) are
mainly due to the fact that in the two hemispheres model, the illuminating
flux is assumed to be constant over the heated regions, whereas the boundary
condition (10) includes an angular variation of this flux, which has a
strongly non-linear effect.

For X-ray luminosities close to the Eddington limit, we predict that the
illumination effect is small, as $x g(x)$ is small for large values of $x$,
and as a significant fraction of the secondary surface might be shielded from
illumination by the accretion disc (Ritter et al., 1996b). However, Eq. (10)
might significantly underestimate the effect of heating, as the heat flux in
the illuminated regions can be negative, and be of the same order of
magnitude as the intrinsic stellar flux. The unilluminated layers have then
to re-radiate this additional flux, which may, in some cases, be sufficient
to also block the intrinsic stellar flux. This is equivalent to having $x
\sim 1$ in these regions which then become very sensitive to illumination. 
This effect, responsible for the short outbursts of mass transfer obtained
by Hameury et al. (1993), is indeed not accounted for by the boundary
condition (10), but depends sensitively on the strength of the coupling
between both hemisphere, i.e. on the circulation timescale which is poorly
known. 

\titlea{Conclusion}

The effects of X-ray illumination of the secondary star in a CV or a LMXB can
be accounted for by a modification of the Stefan-Boltzmann law which we give
in a tabular form. This approximation is quite accurate in CVs, but may lead
to a significant underestimate of the effects of illumination in LMXBs, when
the luminosity is close to the Eddington limit. This modified boundary
condition can be used in bi-polytropic codes, which are much faster than full
stellar codes, and hence allow the exploration of a larger parameter space.

\begref{References}

\ref Frank J., King A.R., Lasota J.P., 1992, \ApJ{385}{L45}
\ref Gontikakis C., Hameury J.M., 1993, \AaA{271}{118}
\ref Hameury J.M., 1991, \AaA{243}{419}
\ref Hameury J.M., King A.R., Lasota J.P., Raison F., 1993, \AaA{277}{81}
\ref Harpaz A., Rappaport S., 1991, \ApJ{383}{739}
\ref Harpaz A., Rappaport S., 1995, \AaA{294}{L49}
\ref King A.R., 1995, In: A. Bianchini, M. Della Valle and M. Orio (eds),
Cataclysmic Variables, Kluwer, p. 523
\ref King A.R., Frank K., Kolb U., Ritter H., 1995, \ApJ{444}{L37}
\ref King A.R., Frank K., Kolb U., Ritter H., 1996, \ApJ{467}{761}
\ref Kolb U., Ritter H., 1992, \AaA{254}{213}
\ref London R., McCray R., Auer L.H., 1981, \ApJ{243}{970}
\ref Podsiadlowski P., 1991 \nat{350}{136}
\ref Ritter H., Zhang Z., Kolb U., 1995, In: A. Bianchini, M. della Valle,
M. Orio (eds) Cataclysmic Variables, Kluwer, p. 479
\ref Ritter H., Zhang Z., Kolb U., 1996a, In: J. Van Paradijs, E.P.J. van den
Heuvel, and E. Kuulkers (eds), IAU Symp. No. 165, Kluwer, p. 65
\ref Ritter H., Zhang Z., Kolb U., 1996b, \AaA{}{to be submitted}
\ref Ritter H., Zhang Z., Hameury J.M., 1996c, In: A. Evans, J. H. Wood (eds),
Cataclysmic Variables and Related Objects, IAU Coll. No. 158, Kluwer, p. 449
\ref Ruderman M., Shaham J., Tavani M., Eichler D., 1989, \ApJ{343}{292}
\ref Tavani M., London R., 1993, \ApJ{410}{281}

\endref
\bye